\documentclass[12pt]{article}

\newcommand{\vex}{{\vec x}}
\newcommand{\be}{\begin{equation}}
\newcommand{\ee}{\end{equation}}
\newcommand{\beqs}{\begin{eqnarray}}
\newcommand{\eeqs}{\end{eqnarray}}

\newcommand{\tr}{{\rm tr}}
\newcommand{\Tr}{{\rm Tr}}
\newcommand{\half}{{1 \over 2}}

\newcommand{\gd}{{g^\dagger}}
\newcommand{\dth}{{\frac{i \delta \theta}{2 \theta^2}}}

\expandafter\ifx\csname mathbbm\endcsname\relax

\else

\fi
\begin{document}
\begin{titlepage}
\begin{flushleft}  
       \hfill                       CCNY-HEP-07/06\\
       \hfill                       December 2007\\
\end{flushleft}
\vspace*{3mm}
\begin{center}
{\LARGE {Noncommutative Wess-Zumino-Witten actions and their Seiberg-Witten invariance}\\}
\vspace*{12mm}
\large {Justo Lopez-Sarri\'on, Alexios P. Polychronakos} \\
\vspace*{5mm}
\large
{\em Physics Department, City College of the CUNY\\
160 Convent Avenue, New York, NY 10031\\
\small justo, alexios@sci.ccny.cuny.edu \/}\\
\vspace*{4mm}
\vspace*{15mm}
\end{center}

\begin{abstract}{

We analyze the noncommutative two-dimensional Wess-Zumino-Witten model and its properties
under Seiberg-Witten transformations in the operator formulation. We prove that the
model is invariant under such transformations even for the noncritical (non chiral)
case, in which the coefficients of the kinetic and Wess-Zumino terms are not related.
The pure Wess-Zumino term represents a singular case in which this transformation
fails to reach a commutative limit. We also discuss potential implications of this
result for bosonization.
}

\end{abstract}

\end{titlepage}

\section{Introduction}

The Wess-Zumino-Witten (WZW) model has been of fundamental importance in various physical contexts.
Generalizations of this model
applying to different systems, such as theories defined on a noncommutative spacetime \cite{MoSha}-\cite{Gheze},
higher dimensional quantum Hall theories \cite{KaNa,PolKM} or higher dimensional bosonization \cite{HiBoz,Dim}, have also appeared.

One of the most interesting applications is in standard bosonization theory \cite{WZW}. 
In its basic form, the WZW model provides a bosonic representation of a free fermionic system (with nonabelian
symmetries) on a two dimensional spacetime. Further, a suitable noncommutative version of the model has been proposed
as an exact bosonization of fermion systems in dimensions higher than two \cite{HiBoz}. As a result, the study of the WZW model
on noncommutative spaces as a tool for bosonization has acquired new interest.

The WZW model on two noncommutative spacetime dimensions (NCWZW) should provide a bosonic description of free fermions on the
same noncommutative spacetime. Such a free fermionic theory, however, does not manifest any observable differences from
the standard (commutative) free fermionic theory, and hence this NCWZW model should be, somehow, equivalent
to the commutative WZW model. In \cite{MoS}, Moreno and Schaposnik showed that there is a transformation that maps
the (critical) NCWZW model to the commutative one. Such mappings, first introduced for gauge theories, are called
Seiberg-Witten (SW) transformations \cite{SW}. Subsequently, the related Chern-Simons action on three noncommutative dimensions
was also shown to be invariant under a SW transformation \cite{GS}.

Noncommutative field theories can be written either in the star-product formulation or in the operator formulation. Although the
two formulations are equivalent, their formalisms often differ substantially in their detail and each gives a different
perspective. The work of \cite{MoS} was in the star-product formulation.  A direct demonstration of the SW invariance of the 
NCWZW action
proves hard to obtain in this formulation and, instead, the invariance of the variation of the action was examined.
It is interesting to examine the same problem in the operator formulation and see if a direct proof of the invariance of the
action itself is attainable.

Another question that naturally arises is whether the criticality of the NCWZW model is crucial for its SW invariance.
In other words, it is interesting to know whether there is a similar SW mapping for the {\it non}-critical NCWZW model, in which the kinetic
and Wess-Zumino terms have different coefficients.

In this work we address the above two issues. We demonstrate that the operator formulation manifests an explicit simplification
for the critical theory and affords a direct proof of the SW invariance of the action (as well as a straightforward proof of
the quantization of the coefficient of the Wess-Zumino term). We further show, in both the star-rpoduct and the operator formulation,
that there is a SW transformation that maps any noncritical NCWZW model to the commutative one, thus
establishing a correspondence for the entire family of models. The only singular point in this family is the ``bare" Wess-Zumino term
(without kinetic term), for which the corresponding transformation becomes singular for vanishing noncommutativity parameter and
thus fails to reach a commutative limit.
 
The plan of this paper is as follows: in section 2 we introduce the star-product formulation, review the work of \cite{MoS}
and generalize it to show the SW invariance of the noncritical NCWZW. In section 3 we review the operator formulation
of the model and of SW transfrmations, apply it to the model at hand and prove the invariance of the action of the
NCWZW model for the noncritical case. Finally, in section 4 we present some conclusions and directions for further research.

\section{The star product formulation}
\subsection{The Wess-Zumino-Witten action}

We shall work on a two-dimensional noncommutative space of Minkowski signature with
light-cone coordinates $\vex = (x,y)$:
\be
ds^2 = \eta_{\mu \nu} dx^\mu dx^\nu = dx dy
\ee
In the star product formulation, noncommutativity amounts to replacing the ordinary product of functions
with the associative but nonlocal and noncommutative star product
\be
f(\vex ) * g(\vex ) = \left. e^{i \frac{\theta}{2} {\vec \partial}_1 \times {\vec \partial}_2}
f(\vex_1 ) g(\vex_2 ) \right|_{\vex_{1,2} = \vex}
\label{starproduct}
\ee
which reproduces the fundamental noncommutativity relation
\be
x * y - y * x = i \theta
\ee
Derivatives and integrals retain their commutative form.

The field variable of the model is an $N \times N$ matrix-valued function of $x,y$ that satisfies a star-unitarity
condition:
\be
g( \vex ) * g(\vex )^\dagger = g( \vex )^\dagger * g (\vex ) = 1
\label{starunitary}
\ee
The kinetic term $K$ and Wess-Zumino term $W$ of the model are the star product transcriptions of the
corresponding commutative terms:
\beqs
K &=& \half \int d^2x \, \tr ( \partial_x g^\dagger * \partial_y g ) = 
-\half \int d^2x \, \tr ( \gd * \partial_x g * \gd * \partial_y g ) \cr
W &=& \frac{1}{4\pi} \int dt d^2x \, \epsilon^{\mu \nu} \tr 
\left( \gd * \partial_t g * \gd *\partial_\mu g * \gd * \partial_\nu g \right)
\eeqs
where $\epsilon^{xy}=1$ is the total antisymmetric tensor.
As in the commutative case, in the Wess-Zumino term the field $g(\vex ,t)$ depends on an additional (commutative)
parameter  $t \in [0,1]$ such that $g(\vex ,0) = 1$, $g(\vex, 1) = g(\vex )$.

The generic action has the form
\be
S = \frac{\lambda}{2\pi} K + k W
\label{Sstar}
\ee
For the particular choices $k=\pm \lambda$ the action reduces to the chiral WZW action. Under a generic 
variation of $g$ (independent of $\theta$) this action's change is $t$-independent and reads
\be
\delta S = \frac{1}{4\pi}(\lambda \eta^{\mu \nu} + k \epsilon^{\mu \nu})
\int d^2x\,\tr\,
\left[\partial_\mu (\gd * \delta g) * \gd * \partial_\nu g)\right]
\label{deltaSstar}
\ee
The resulting equation of motion is the star-analog of the corresponding commutative one:
\be
A^{\mu \nu} \partial_\mu J_\nu = 0
\label{starEOM}
\ee
where we defined the constant matrix $A^{\mu\nu}$ and the antihermitian currents $J_\mu$
\be
A^{\mu \nu} = \lambda \eta^{\mu \nu} + k \epsilon^{\mu \nu} ~,~~~~ J_\mu = \gd * \partial_\mu g
\ee
For $k = \pm \lambda$ the equation of motion becomes obviously chiral, $\partial_x J_y =0$ or $\partial_y J_x = 0$.

\subsection{Seiberg-Witten transformation}

Seiberg-Witten transformations map an action to a (generically different) action on a space of
different noncommutativity parameter. Such an infinitesimal transformation in the star product formulation
(denoted SW$_*$ to distinguish it from the corresponding form of the transformation in the operator formulation)
induces a specific variation in the field $g$ (denoted $\Delta_* g$) {\it and} a variation in $\theta$,
$\delta \theta$. As a result, star products of fields change both because of the change in the fields and the
change in the definition of the star product. From (\ref{starproduct}) we deduce that
\be
\Delta_* (A * B) = \Delta_* A * B + A * \Delta_* B + i\frac{\delta \theta}{2} \epsilon^{\mu \nu} \partial_\mu A * \partial_\nu B
\label{varstar}
\ee
for any $A,B$.

The SW$_*$ variation $\Delta_* g$ must obey a consistency condition arising from the star-unitarity of $g$. Varying
(\ref{starunitary}) we obtain
\be
\Delta_* (\gd * g) = \Delta_* \gd * g + \gd * \Delta_* g + 
i \frac{\delta \theta}{2} \left(\partial_x \gd * \partial_y g - \partial_y \gd * \partial_x g \right)=0
\ee
Upon using $\partial_\mu \gd = - \gd * \partial_\mu g * \gd$, the above can also be written as
\be
\Omega + \Omega^\dagger = \frac{i}{2} ( J_x * J_y - J_y * J_x )
\label{omom}
\ee
where we defined the SW$_*$ variation matrix
\be
\Omega = \gd * \frac{\Delta_* g}{\delta \theta}
\ee
For an ordinary transformation (with $\delta \theta =0$) the variation matrix $\gd * \delta g$ is antihermitian.
When $\theta$ varies as well, however, equation (\ref{omom}) shows that $\Omega$ acquires a hermitian part proportional
to the star-commutator of the two currents. Any form for $\Omega$ consisting of an arbitrary antihermitian part plus the
fixed hermitian part would be allowed. The above equation, however, suggests two obvious ``minimal" forms for $\Omega$:
since $(i J_x * J_y )^\dagger = -i J_y * J_x$, we can choose
\be
\Omega_+ = \frac{i}{2} J_x * J_y ~~~{\rm or}~~~~
\Omega_- = - \frac{i}{2} J_y * J_x
\label{omin}
\ee
The first of these is exactly the choice made in the work of Moreno and Schaposnik \cite{MoS}.
They showed that the noncommutative version
of the critical WZW model is invariant under the above particular Seiberg-Witten transformation.
To do that, they had to use the equations of motion. Indeed, they proved that this transformation left invariant
not the action itself but the variation of the action with respect to a generic variation of the field $g$
without changing the noncommutative parameter.

To be more concrete, they proposed as Seiberg-Witten transformation,
\beqs
\frac{dg}{d\theta} &=& -\frac{i}2 \partial_x g * \partial_y \gd * g \cr
\frac{dg^\dag}{d\theta} &=& \frac{i}2 \gd * \partial_x g * \partial_y \gd
\eeqs
Upon use of $\partial_\mu \gd = - \gd * \partial_\mu g * \gd$ and $\Delta_* g = \frac{dg}{d\theta} \delta\theta$,
it is clear that the above equations correspond to the first of the ``minimal" choices in (\ref{omin}).
Then, they proceeded to prove that this transformation leaves invariant the noncommutative version of the 
WZW model with critical coefficient $k=-\lambda$.

As we shall demonstrate, however, a more general variation of the same type, consistent with the
relation (\ref{omom}), leaves the equations of motion of the general action invariant. This transformation is
written in terms of an arbitrary real parameter $\beta$ as
\beqs
\Omega = g^\dag * \frac{dg}{d\theta} &=& -\frac{i}{4}\epsilon^{\mu\nu}
 \partial_\mu g^\dag * \partial_\nu g + 
\frac{i(2\beta-1)}{4}\eta^{\mu\nu}\partial_\mu g^\dag * \partial_\nu g \cr
&\equiv& \frac{i}{2} B^{\mu\nu} \partial_\mu g^\dag * \partial_\nu g
\label{stargeneral}
\eeqs
where we defined the matrix $B$ as
\be
B^{\mu\nu} = -\frac{1}{2}\epsilon^{\mu\nu} - \frac{2\beta -1}{2}\eta^{\mu\nu}
\ee
Note that the case $\beta=0$ is the particular case used in the paper \cite{MoS}. The case 
$\beta=1$ corresponds to interchaging $g$ and $g^\dag$ and will leave invariant the critical WZW model
of opposite chirality. Hence, our new transformation is a linear combination of two chiral transformations, {\it i.e.},
\be
\Omega =(1-\beta) \Omega_+ + \beta \Omega_-
\ee

The question that arises at this point is what action is preserved by the above transformation. A direct evaluation
of the transformation of the general action (\ref{Sstar}) in the star-product formulation leads to some unwanted terms
that require the use of highly nontrivial identities in order to cancel. 
We shall, instead, follow the method of \cite{MoS} and examine the invariance of the variation of the action.
In the operator formalism, as we shall see in the next section, the invariance of the action can be
directly proven.

Following \cite{MoS} we can perform a generic $\theta-$independent transfomation $\delta g$ for which
\be
\frac{d}{d\theta} \delta g = \delta \left( \frac{dg}{d\theta} \right)
\ee
Defining
\be
\omega = g^\dag * \delta g ~,~~~~ \Omega = \gd * \frac{dg}{d\theta}
\ee
we can directly check, upon using (\ref{varstar}), that
\be
\frac{d\omega}{d\theta} = \left[\omega, \Omega \right]_* +
\delta \Omega - 
\frac{i}{2}\epsilon^{\mu\nu}g^\dag * \partial_\mu g * \partial_\nu\omega
\ee
Substituting the specific variation (\ref{stargeneral}) with respect to $\theta$ we obtain,
\be
\frac{d\omega}{d\theta} = \frac{i}{2}B^{\mu\nu} (g^\dag * \partial_\mu g) * \partial_\nu\omega + 
\frac{i}{2}B^{\mu\nu}\partial_\nu\omega *  (g^\dag * \partial_\mu g)
\ee
For the particular case where $\delta g = \partial_\beta g$ we obtain the SW$_*$ variation of the current $J_\beta$
\be
\frac{dJ_\beta}{d\theta} = \frac{i}{2} B^{\mu\nu} J_\mu * \partial_\nu J_\beta + 
\frac{i}{2} B^{\mu\nu}\partial_\nu J_\beta * J_\mu
\label{swcurrent}
\ee

Now consider the general action (\ref{Sstar}) whose general variation $\delta S$ is given by (\ref{deltaSstar}).
By applying the above equations for the SW$_*$ variation of $\omega = \gd \delta g$ and $J_\beta$, we get
for the SW$_*$ variation of $\delta S$
\begin{eqnarray}
\frac{d\delta S_\kappa}{d\theta} &=& \frac{i}{8\pi} A^{\alpha \beta} \int\,d^2x\,{\rm tr}\,\Bigl[ B^{\mu\nu}
\left(\partial_\alpha\omega * J_\mu * \partial_\nu J_\beta - 
J_\mu * \partial_\nu\omega *  \partial_\alpha  J_\beta \right)\nonumber\\
&& ~~~~~~~~~~~~~~~~~~+B^{\mu\nu}\left(\partial_\alpha\omega * \partial_\nu J_\beta *  J_\mu - 
\partial_\nu \omega *  J_\mu * \partial_\alpha J_\beta \right)\Bigr]\nonumber\\
&=& \frac{i}{8\pi}\int d^2x\,{\rm tr}\left[(A^{\alpha\beta} B^{\mu\nu} - 
A^{\nu\beta} B^{\mu\alpha})\partial_\alpha\omega *  J_\mu * \partial_\nu J_\beta\right.\nonumber\\
&+&\left. (A^{\alpha\beta} B^{\nu\mu} - 
A^{\mu\beta}B^{\nu\alpha})\partial_\alpha\omega * \partial_\mu J_\beta *  J_\nu\right]\nonumber\\
&=&\frac{i}{8\pi}\int d^2x\,{\rm tr}\left[(A^{\alpha\nu} B^{\mu\beta} - 
A^{\beta\nu} B^{\mu\alpha})\partial_\alpha\omega *  \left\{J_\mu,\partial_\beta J_\nu\right\}_* \right]
\nonumber\\
&=&-\frac{i}{8\pi}\int d^2x\,{\rm tr}\left[(A^{\alpha\nu} B^{\mu\beta} - A^{\beta\nu} B^{\mu\alpha})
\omega *  \left\{\partial_\alpha J_\mu,\partial_\beta J_\nu\right\}_* \right]\label{final}
\end{eqnarray}

The matrix factor in the above equation (\ref{final}) is antisymmetric under exchange of the indices $\alpha$ and
$\beta$ and is, hence, proportional to $\epsilon^{\alpha\beta}$. Moreover, because the integrand is symmetric with
respect to the exchange of indices $(\alpha,\mu)\leftrightarrow (\beta,\nu)$,  
the tensorial structure of this equation has to be of the form
\be
\frac{d\delta S_\kappa}{d\theta}= \frac{iC}{8\pi}\epsilon^{\alpha\beta}\epsilon^{\mu\nu}\int d^2x\,{\rm tr}
\left[\omega *  \left\{\partial_\alpha J_\mu,\partial_\beta J_\nu\right\}_ * \right]
\ee
where the coefficient $C$ is given by
\be
C =\frac{1}2\epsilon_{\alpha\beta}\epsilon_{\mu\nu}A^{\beta\nu}B^{\mu\alpha}= (2\beta -1)\lambda - k
\ee
We see, therefore, that the condition $C=0$, or
\be
k = (2\beta - 1) \lambda
\ee
ensures the invariance of $\delta S$ under the SW$_*$ transformation (\ref{stargeneral}).

The above derivation implies that the equation of motion is invariant under the SW transformation.
It is worth checking the invariance of the equation of motion explicitly.
To do so, we take the variation of the left-hand side of (\ref{starEOM}) $E = A^{\mu\nu}\partial_\mu J_\nu$
and use equation (\ref{swcurrent}) for the transformation of the current, 
\begin{eqnarray}
\frac{d E}{d\theta} &=& A^{\alpha\beta}\partial_\alpha \frac{dJ_\beta}{d\theta} 
= -\frac{i}2 A^{\alpha\beta}B^{\mu\nu}\partial_\alpha \left\{J_\mu,\partial_\nu J_\beta\right\}_* \nonumber\\
&=&-\frac{i}2A^{\alpha\beta}B^{\mu\nu}\left\{\partial_\alpha J_\mu,\partial_\nu J_\beta\right\}_* 
- \frac{i}2 B^{\mu\nu}\left\{J_\mu,\partial_\nu E\right\}_* \nonumber\\
&=&-\frac{i}2  C \epsilon^{\alpha\nu}\epsilon^{\beta\mu}\left\{\partial_\alpha J_\mu,\partial_\nu J_\beta\right\}_*  
-\frac{i}2 A^{\nu\beta}B^{\mu\alpha} \left\{ \partial_\alpha J_\mu,\partial_\nu J_\beta\right\}_*\nonumber\\
&& -\frac{i}2 B^{\mu\nu}\left\{J_\mu,\partial_\nu E\right\}_*
\end{eqnarray}
where in the last step we have added and substracted the last term in the third  line, and we have made use of the same 
arguments about the symmetries of the indices as above. Imposing the condition $C=0$ by adjusting $\beta$, we end up with the result
\be
\frac{d E}{d\theta} =-\frac{i}2B^{\alpha\mu} \left\{ \partial_\alpha J_\mu,E\right\}_* - 
\frac{i}2 B^{\mu\nu}\left\{J_\mu,\partial_\nu E\right\}_*
\ee
When the equation of motion $E=0$ is satisfied we see that the above variation vanishes, which proves that the equation of motion
is left invariant by the given $\theta-$transformation.

The above general result reproduces the special chiral cases of $\beta=0$, corresponding to the critical coefficient $k=-\lambda$,
and $\beta = 1$, corresponding to the critical coefficient $k = \lambda$. For other values of $\beta$ it
extends the SW invariance to actions with general $k$ and $\lambda$. We postpone a further discussion
of this transformation until after we prove the full invariance of the action in the operator
formulation.

\section{Operator formulation}
\subsection{Operator form of the action}

In the previous section we worked with a transformation of the variation of the action,
rather than the action itself.
The transformation properties of the action become much more transparent in the
operator formulation. In this section we present the operator form of the WZW
action and the corresponding theta-rescaling SW transformations.

In the operator formulation, the coordinates $x,y$ become operators satisfying
\be
[x,y]=i\theta
\label{xycomm}
\ee
and act on $N$ copies of the irreducible representation of the above Heisenberg
commutation relation. Fields become arbitrary operators on this space and can be
parametrized as $N \times N$ matrices with elements operators expressible in terms
of $x$ and $y$ (see, e.g., \cite{GN,AGW,PCS}). In particular, the $U(N)$ field $g$
becomes simply a unitary operator, satisfying
\be
g g^\dagger = g^\dagger g = 1
\ee
Derivatives of any operator $F$ are obtained throught the adjoin action of $x,y$ as
\be
\partial_x F = \frac{i}{\theta} [y,F] ~,~~~ \partial_y f = -\frac{i}{\theta} [x,F]
\ee
and combined integration over $x,y$ and ($N \times N$) matrix trace become trace over
the full Hilbert space:
\be
\int d^2 x \, \tr F = 2\pi\theta \, \Tr F
\ee

It is important for our considerations to examine the properties of the above trace
under cyclic permutations of operators. It is clear that cyclicity of trace is in general
not valid, as, e.g., applied in the fundamental relation (\ref{xycomm}) it would give
$0 = i\theta \Tr 1$, an obvious absurdity. The problem is the same as integration by
parts in the star-product formulation: the integral of a total derivative (corresponding
to the trace of a commutator) amounts to boundary terms at infinity that can be nonzero.

To properly use cyclicity of trace, we note that it is still valid for traces of the
form
\be
\Tr (C f) = \Tr (f C)
\ee
where $C$ is a {\it compact} operator and $f$ is arbitrary. In particular, if $C$ is a finite
matrix in the oscillator basis $|n\rangle$ then the above is true ($C$ is the analog of a
function of compact support). We note, further, that operators of the form $p C q$, where
$C$ is compact as above and $p,q$ are arbitrary polynomials in $x,y$ of finite order, are
also compact in the oscillator basis. This is due to the fact that $x$ and $y$ are ``local"
operators; that is, their action on $|n\rangle$ produces only a finite number of nearby
states $| n\pm 1 \rangle$ and thus $pCq$ is also a finite matrix. This is all we need
for our considerations.

For the present model, we shall make the usual assumption that $g$ falls off to unity fast
enough at infinity; that is, it can be written as $g = 1 + G$, with $G$ compact.
This immediately implies that, for any $f$,
\be
\Tr (g f) = \Tr (f g)
\ee
The compactness of $pGq$, on the other hand, also implies
\be
\Tr (pgq) = \Tr (pq + pGq) = \Tr (pq + Gqp) = \Tr (gqp + [p,q])
\ee
Similar relations hold true also for $g^\dagger$. The above, generalized to operators of the form
$A = p g q g^\dagger r \dots $, implies
\be
\Tr [ A(g) , B(g) ] = \Tr [ A(1) , B(1) ]
\label{TrAB}
\ee
In other words, we {\it can} use cyclicity of trace in expressions containing
$g$ and $g^\dagger$, as long as we correct them by adding the commutator of the ``bare" quantities
obtained by putting $g = g^\dagger = 1$.

Properly speaking, expressions like the above may not have a finite trace.
The validity of the cyclicity of trace is related to the validity of exchanging
the order of (infinite) summations in expressions that are, otherwise, finite. The above manipulations
mean that we can substitute $pgq$ with $gqp+[p,q]$ in all expressions inside a trace provided the
overall trace is finite. As an example, we have
\be
\Tr (x g y g^\dagger - xy) = \Tr (y g^\dagger x g + [x,y] -xy) = \Tr (y g^\dagger x g -yx)
\ee

We are now ready to examine the operator form of an the action $S = \frac{\lambda}{2\pi} K + k W$.
Under the above conditions, the kinetic energy term of the action is easily seen to take the form
\be
K = \half \int d^2x \, \tr ( \partial_x g^\dagger \partial_y g ) =
\frac{\pi}{\theta} \Tr \left( g x \gd y + \gd x g y - 2xy \right)
\ee
We note that $K$ satisfies, as expected,
\be
K[g] = K[\gd ]
\ee
The Wess-Zumino term, on the other hand, is
\beqs
W &=& \frac{1}{4\pi} \int dt d^2x \, \tr \left( \gd {\dot g} [ \gd \partial_x g , \gd \partial_y g ] \right)
\cr
&=& \frac{1}{2\theta} \int dt \Tr \left[ \gd {\dot g} \left( y \gd x g - \gd x g y - x \gd y g + \gd y g x
+ 2i\theta \right) \right]
\eeqs
The first four terms, using the fact that ${\dot g} = {\dot G}$ is a compact operator and cyclicity of trace
relations, are actually a total derivative in $t$. Integrating them gives a purely two-dimensional
term and we obtain
\be
W = i \int dt \Tr ( \gd  {\dot g} ) + \frac{1}{2\theta} \Tr (x g y \gd - x \gd y g )
\ee
We observe that the above is almost entirely local, the only remnant of the auxiliary variable
$t$ being in the first, Kostant-Kirillov-Souriau type term for $g$. Note, further, that the
coefficient of this term is $\theta$-independent and the ``$K$-matrix" multiplying the one-form
$\gd {\dot g}$ is the identity, essentially isolating only the abelian part of $g$.
Standard arguments for the quantization of the KKS term imply that its coefficient $k$ must be an integer,
therefore recovering the quantization of the coefficient of the WZ term in the noncommutative
setting. (Similar arguments apply to the noncommutative Chern-Simons term \cite{NP,Klee}.)
As expected, $W$ satisfies
\be
W [ \gd ] = - W[g]
\ee

It is interesting to construct the operator expressions for the chiral WZW actions 
\be
S_k = |k| S_\pm = |k| \left( \frac{1}{2\pi} K \pm W \right) 
\ee
corresponding to tuning the
coefficient of the kinetic term to be $\pm 1/2\pi$ times the coefficient of WZ term, depending on
the sign of $k$. 
Precisely for these two values of the coupling constants, we observe that a partial cancellation
happens between terms in $K$ and $W$ and the form of the action simplifies considerably. We obtain for
the two fundamental chiral actions $S_\pm$ (corresponding to $k = \pm 1$)
\beqs
S_+ &=& i\int dt \Tr ( \gd  {\dot g} ) + \frac{1}{\theta} \Tr (xgy \gd - xy) \cr
S_- &=& -i\int dt \Tr ( \gd  {\dot g} ) + \frac{1}{\theta} \Tr (x \gd yg - xy)
\eeqs
The equations of motion derived from the above actions are obtained by calculating the variation
of the action under a general variation of $g$. Since $g=1+G$, $\delta g = \delta G$ is compact
and we can freely use cyclicity of trace. We obtain
\beqs
\delta S_+ [g] &=& \frac{1}{\theta} \Tr \left[ E(g) \gd \delta g \right] \cr
\delta S_- [g] &=& -\frac{1}{\theta} \Tr \left[ E(\gd ) \delta g \gd \right] = \delta S_+ [\gd ]
\eeqs
In the above we defined the operator quantity
\be
E(g) = y\gd xg - \gd x g y + i\theta = [y , \gd x g - x] = \gd [ g y \gd - y , x ] g
\label{Eg}
\ee
The equation of motion for the action $S_+$ ($S_-$ resp.) is given by setting the quantity $E(g)$
($E(\gd )$ resp.) to zero:
\begin{eqnarray}
\delta S_+ = 0 &\to ~ [y, \gd x g - x ] = 0  &{\rm or} ~~[x, g y \gd - y ] = 0 \cr
\delta S_- = 0 &\to ~ [y, g x \gd - x ] = 0  &{\rm or} ~~[x, \gd y g - y ] = 0 
\end{eqnarray}
The chiral form of these equations is obvious.

The general non-chiral action can always be written as a linear combination of the above two
chiral actions, as
\be
S = \frac{\lambda}{2\pi} K + k W = \frac{\lambda + k}{2} S_+ + \frac{\lambda - k}{2} S_-
\label{generalS}
\ee
We shall adopt a parametrization of $k$ in terms of an interpolating parameter $\alpha$,
\be
k = \lambda (2\alpha -1)
\ee
bringing the general action to the form
\be
S = \lambda \left[ \alpha \, S_+ + (1-\alpha ) \, S_- \right]
\ee
The variation of $S$ becomes
\be
\delta S = \lambda \Tr \left[ \alpha \, E(g) \gd \delta g - (1-\alpha ) \, E(\gd ) \delta g \gd \right]
\label{varS}
\ee
with a corresponding expression for the equation of motion, which is not chiral any more.

\subsection{Operator form of the SW transformation}

The operator form of the SW transformation for gauge fields had been worked out in \cite{KS,APSW}.
The main point of difference from the star-product formulation, as was detailed in \cite{APSW}, is
that the operator's variation has two sources: one is the change of the `symbol' of the operator,
represented by the transformation of the corresponding function in the star-product
formulation, and another is the change of the underlying operators $x,y$ in terms of which this
function is defined. The overall change of any operator $A$ is
\be
\Delta A = \Delta_* A + \frac{i \delta \theta}{2\theta^2} \left(i\theta A -xAy + yAx \right)
\label{opSW}
\ee
where $\Delta_* A$ stands for the change in the operator $A$ due to the change in its symbol
(the commutative function $A(x,y)$ representing $A$ in the star-product formalism).
In the above and hereon we adopt the notation $\delta A$ for an arbitrary variation of $A$ and
$\Delta A$ for the specific variation corresponding to the SW transformation.

An important point about the above transformation is that it is a true variation; that is
\be
\Delta (A B) = \Delta A \, B + A \, \Delta B
\ee
This should be contrasted to the star-product SW transformation (SW$_*$ for short), where 
$\Delta_* (A * B)$ contains extra terms due to the $\theta$-dependence in the definition of
the star product, as detailed in the section on the star-product formulation.

The extra term in the transformation (\ref{opSW}) implies that even functions invariant under SW$_*$
do transform in the operator formulation. In particular, the coordinates themselves (invariant under
SW$_*$) transform as
\be
\Delta x = \frac{\delta \theta}{2\theta} x ~,~~~ \Delta y = \frac{\delta \theta}{2\theta} y
\ee
Similarly, monomials in the coordinates of the form $x^m y^n$ (in any ordering) transform as
\be
\Delta (x^m y^n ) = \delta \theta \frac{m+n}{2\theta} \, x^m y^n
\ee
The above monomials would map to star-product functions of the form $x^m y^m + O(\theta)$
due to Weyl reordering (or, equivalently, due to the star products between the single factors
of $x$ and $y$). Such expressions do transform under SW$_*$ due to the appearance of explicit
$\theta$ factors, and this is neatly captured by the additional term in (\ref{opSW}) above.

Importantly, combinations of the form $\theta^{-n} x^m y^{2n-m}$ are invariant:
\be
\Delta \left( \frac{1}{\theta^n} x^m y^{2n-m} \right) = 0
\ee
Consequently, operators of the form $\theta^{-1} A \, x \, B \, y$ transform only due
to $A$ and $B$:
\be
\Delta \left( \frac{1}{\theta} A \, x \, B \, y \right) = \frac{1}{\theta} 
\Delta A \, x \, B \, y + \frac{1}{\theta} A \, x \, \Delta B \, y
\ee
This will be relevant for the WZW action.

We now have all the ingredients to write the operator version of the SW transformation.
The chiral actions $S_+$ and $S_-$ transform differently in the star-product formulation.
The two transformations, denoted $\Delta_{+*} g$ and $\Delta_{-*} g$, respectively, are
\beqs
\gd * \Delta_{+*} g &=& \frac{i \delta \theta}{2} J_x * J_y \cr
\gd * \Delta_{-*} g &=& -\frac{i \delta \theta}{2} J_y * J_x
\eeqs
To find the operator version of the above we omit star products, write the
operator version of the currents
\beqs
J_x &=& \frac{i}{\theta} \gd [ y , g ] \cr
J_y &=& -\frac{i}{\theta} \gd [ x , g ]
\eeqs
and apply formula (\ref{opSW}) above. The result is
\beqs
\Delta_+ g &=& \dth \, g D(g) \cr
\Delta_- g &=& -\dth \, D(\gd ) g
\eeqs
where we defined the operator quantity
\be
D(g) = \gd xy g + yx - \gd x g y - y \gd x g
\label{Dg}
\ee
We observe that $D(g)$ is hermitian and thus the above expressions satisfy
\be
( \Delta_\pm g )^\dagger = - \gd \, \Delta_\pm g \, \gd = \Delta_\pm (\gd )
\ee
and so they preserve the unitarity of $g$.
The two transformations are also chirally related in the sense that $(\gd \Delta_+ g)(x,y)
= -(\Delta_- g \, \gd )(y,x)$, as indicated by their expressions in terms of $D(g)$.

The general, non-chiral action $S$ is not invariant under any of the transformations above.
As in the star-product formulation, we shall try as an ansatz for the general SW transformation a 
linear combination of the above two transformations:
\be
\Delta g = \beta \Delta_+ g + {\bar \beta} \Delta_- g = 
\left[ \beta g D(g) - {\bar \beta} D(\gd ) g \right] \delta \theta ~,~~~ \beta + {\bar \beta} =1
\label{genD}
\ee
The fact that the two two coefficients $\beta$ and ${\bar \beta} = 1-\beta$ in (\ref{genD})
sum to 1 is crucial, as it ensures that the generalized transformation still has a good commutative
limit. Indeed, it maps to the linear combination of the two chiral transformations in
the star-product formulation:
\be
\Delta_* g = \beta \Delta_{+*} g + {\bar \beta} \Delta_{-*} g
\ee
If, instead, we had chosen two arbitrary coefficients $\beta,{\bar \beta}$, then the SW$_*$
transformation would be
\be
\Delta_* g = \beta \Delta_{+*} g + {\bar \beta} \Delta_{-*} g + \frac{1-\beta-{\bar \beta}}{2\theta}
(x \partial_x g + y \partial_y g) \delta \theta
\ee
which is obviously singular at $\theta=0$.

\subsection{Proof of invariance of the action}

We are now ready to examine the transformation properties of the general WZW action under
SW transformations.

A key observation is that the operator form of the actions $S_\pm$, as well as the general
action (\ref{generalS}), contains a part that is both $\theta$- and $x,y$- independent (the first,
KKS term) plus terms of the form $\theta^{-1} \Tr ( A x B y)$ with $A,B$ operators that do not depend
on $\theta$, $x$ or $y$ (namely $g$, $\gd$ and $1$).
As explained in the previous subsection, such terms respond to SW transformations
only through the variation of $A,B$ ($\theta^{-1} xy$ is SW invariant). Therefore, the full SW variation of
the action can be obtained by using the variation of the action $\delta S$ under a general $\delta g$
and putting $\delta g = \Delta g$. This is in contradistinction to the SW$_*$ care, where the variation
of the action picks up extra contributions from the star products.

Using formula (\ref{genD}) for the SW transformation inside the general variation formula (\ref{varS})
and writing ${\bar \alpha} = 1-\alpha$ for brevity we obtain
\beqs
\Delta S = \lambda \dth \, \Tr \Bigl[ &\alpha \beta E(g) D(g) - \alpha {\bar \beta} E(g) \gd D(\gd ) g \cr
&+ {\bar \alpha}{\bar \beta} E(\gd ) D(\gd ) - {\bar \alpha} \beta E(\gd ) g D(g) \gd \Bigr]
\label{DS}
\eeqs
The top left and bottom left term in the above expression correspond to the variations
$\Delta_+ S_+$ ($\Delta_- S_-$ resp.), as can be seen by putting $\alpha=\beta=1$
($\alpha=\beta=0$ resp.) while the two terms to the right are ``mixed" nonchiral terms.

Upon using formulae (\ref{Eg}) and (\ref{Dg}) for $E(g)$ and $D(g)$, the evaluation of
$\Delta S$ can be performed by a brute force calculation. It is instructive, nevertheless, to perforn this
calculation in steps that reduce the overall complexity and reveal the basic interplay between the
SW transformation and the WZW action.

We first deal with the term $\Tr E(g) D(g)$. We notice that it can be written as
\be
\Tr [ E(g) D(g) ] = \Tr \left\{ [y , \gd x g - x] \left( -\{y , \gd x g - x \} + [ \gd xy , g ] \right) \right\}
\ee
The operator $\gd x g - x \sim J_y$ is compact, so we can freely use cyclicity of trace in the terms it appears.
The first term in the above expression, therefore, being of the form $\Tr [y,J_y ] \{ y,J_y \}$, vanishes.
The remaining term is, upon expansion and cyclic permutation of $g$,
\beqs
\Tr [ E(g) D(g) ] &=& \Tr \left( gy\gd xxy - y\gd xgxy - xgy\gd xy + \gd xgyxy \right) \cr
&=& \Tr \left( gy\gd xyx + i\theta gy\gd x - y\gd xgyx - i\theta y\gd xg - xgy\gd xy +\gd xgyxy \right) \cr
&=& \Tr \left( [gy\gd xy,x] + [\gd xgyx,y] \right)
\eeqs
According to formula (\ref{TrAB}), we can replace the above commutators with the bare $g=1$ expressions
\be
\Tr [E(g) D(g) ] = \Tr \left( [yxy,x] + [xyx,y] \right) = \Tr ( 0 ) = 0
\ee
The above shows that $\Delta_+ S_+ =0$ and (upon putting $g \to \gd$) $\Delta_- S_- =0$, thus
establishing the SW invariance of the chiral WZW action.

To show the invariance of the general action $S$
we must deal with the remainig two mixed terms. It is enough to calculate $\Tr [ E(g) \gd D(\gd ) g ]$,
the other one being obtained by putting $g \to \gd$. In this term no obvious cancellations happen
and it reads
\be
\Tr [ E(g) \gd D(\gd ) g ] = \Tr \left[ ( y\gd xg - \gd x g y + i\theta )
( \gd xy g + yx - \gd y g x - x \gd y g ) \right]
\ee
Expanding the above expression and using cyclicity of trace, correcting with the commutator
of bare quantities as in (\ref{TrAB}), we obtain after a fair amount of algebra
\beqs
\Tr [ E(g) \gd D(\gd ) g ] &=& \Tr \Bigl[ \gd xgy\gd ygx + 2i\theta xgy\gd 
- gx\gd ygy\gd x - 2i\theta x\gd yg \Bigr] \cr
&=& - \Tr \bigl[ E(\gd) g D(g) \gd \bigr]
\eeqs
The total variation (\ref{DS}), therefore, becomes
\be
\Delta S = \lambda \dth \, ({\bar \alpha} \beta - \alpha {\bar \beta})  \,
\Tr \left[ E(g) \gd D(\gd ) g \right]
\ee
We see that upon choosing
\be
{\bar \alpha} \beta - \alpha {\bar \beta} = 0 ~,~~~ {\rm or} ~~~ \alpha = \beta
\label{conditionab}
\ee
The total variation vanishes. We have therefore proved that the transformation
\be
\Delta g = \alpha \Delta_+ g + (1-\alpha ) \Delta_- g = 
\left[ \alpha g D(g) - (1-\alpha) D(\gd ) g \right] \delta \theta 
\ee
leaves the general action $S(\lambda,\alpha)$ invariant.

\subsection{A few special cases}

It is interesting to consider a few special cases of the above. The chiral limits
$\alpha=1$ and $\alpha=0$ have already been mentioned and correspond to the
known SW invariance of the chiral WZW action.

Another special case is the pure nonlinear sigma model, without the Wess-Zumino term,
obtained at $\alpha = \half$. The form of the transformation becomes
\be
\gd \Delta_{_K} g = \frac{i \delta \theta}{4\theta^2} \left( \{x,\gd yg\} - \{y,\gd xg\} \right)
\ee
and it leaves invariant the kinetic term $K$ alone.

The case of the pure Wess-Zumino term is interesting: it corresponds to $\alpha = \pm \infty$
and appears singular. This is, of course, an artifact of the parametrization in terms of
$\alpha = (k/\lambda +1)/2$ which becomes singular at $\lambda =0$. Still, the corresponding
SW transformation is singular, as it does require $\beta = \alpha = \pm \infty$. There is 
{\it no} SW transformation with a good commutative limit that leaves the pure Wess-Zumino
term $W$ invariant.

Nevertheless, we can write nontrivial SW transformations that leave $W$ invariant, as long
as we do not insist on a regular commutative limit. The condition (\ref{conditionab}) for
invariance of this term really reads
\be
\lambda ({\bar \alpha} \beta - \alpha {\bar \beta} ) = 0 ~~~{\rm or}~~~ 
(\lambda - k) \beta - (\lambda + k) {\bar \beta} = 0
\ee
and for $\lambda = 0$ is satisfied for $\beta + {\bar \beta} = 0$.
Choosing any nonzero value of $\beta$ leads to
\be
\gd \Delta_{_W} g = \beta \dth \{ \gd x g - x , \gd y g - y \}
\ee
The corresponding SW$_*$ transformation is
\be
\gd * \Delta_* g = \frac{i \beta \delta \theta}{2} \left[ J_x * \left( J_y + 
\frac{x}{i\theta \beta} \right)
+ J_y * \left( J_x + \frac{y}{i\theta \beta} \right) \right]
\ee
which diverges at $\theta =0$ but remains well-defined for any nonzero $\theta$.

\section{Conclusions and discussion}

In conclusion, we have demonstrated the invariance of the WZW action under $\theta$-rescaling
SW transformations, for arbitrary noncritical coupling constants for the kinetic and Wess-Zumino
terms. The derivation is most straightforward in the operator formulation, which also exhibits an
explicit cancellation between terms for critical coupling.

The above results are likely to be relevant to questions related to bosonization. In two
dimensions, the story for the critical commutative model is more or less complete, as it is known
to describe free fermions. For higher dimensions, however, the situation is not so clear. According
to recent work, a noncommutative Kac-Moody type algebra describing the edge dynamics of the Fermi sea
is the correct framework for higher dimensional bosonization \cite{HiBoz}. The lagrangian setting for
such a theory would be a WZW type of action. One possible derivation of such an action would be to start
with a full description of the phase space dynamics of the fermionic many-body system in terms of a
KKS action and suitably reduce it to boundary dynamics \cite{NCF}. Such a reduction is very much like a partial
SW transformation that renders two of the coordinates commutative (the ``radial" coordinate of the
phase space Fermi sea and its conjugate variable) while leaving the rest coordinates noncommutative.
The SW invariance analyzed in the present paper is, therefore, a step in this direction. The issues
that need to be addressed are, first, the presence of the remaining noncommutative dimensions and,
second, the different geometry and topology of the spaces: in the case treated here, the two-dimensional
space is noncommutative while the auxilliary variable $t$ is commutative, while in the bosonization case
the noncommutative space consists of the auxiliary variable and one of the remaining coordinates.
The appearance of the NCWZW action in the context of the NC Sine-Gordon model and its possible
connection to a Thirring-like model \cite{Olaf} are further indications of its relation to bosonization.

A related issue is the physical meaning of the above SW transformation. For noncommutative gauge theories
a hydrodynamic interpretation is possible, in which the SW transformation amounts to the transition from
(comoving) Lagrange coodrinates to (space-fixed) Euler variables \cite {JP}. It would be interesting to
attribute a similar or analogous interpretation to the SW transformation for the WZW model.

Finally, another interesting issue that emerges in the operator formulation is the fact that the SW
transformation is a true variation of the theory: unlike the star-product formulation, it takes the
form of a specific infinitesimal change of the field that leaves the action invariant. It should, therefore,
lead to the appearance of the noncommutative analog of a conservation law. The form and significance of this law,
as well as questions related to bosonization, are under investigation.

\vskip 0.2in
\noindent
{\it \underline {Acknowledgements}:} 
A.P.P. acknowledges the support of the National Science Foundation under grants
PHY-0353301 and PHY-0555620, and of the CUNY Research Foundation under grant PSC-CUNY-69145-0038;
J.L.-S. acknowledges the support of the Fulbright Foundation under grant MEC/Fulbright FU2006-0469.

\end{document}